\documentclass[aip,pop,preprint,groupedaddress]{revtex4-1}
\usepackage{amsmath}
\usepackage{subfigure}
\usepackage{graphicx}
\usepackage{color}

\begin{document}



\title{Investigating the effects of electron bounce-cyclotron resonance on plasma dynamics in capacitive discharges operated in the presence of a weak transverse magnetic field}



\author{Sarveshwar Sharma,} %
\email[Corresponding author:]{ sarvesh@ipr.res.in, sarvsarvesh@gmail.com}%
\author{Sudip Sengupta, Abhijit Sen}%
\affiliation{Institute for Plasma Research, Bhat, Gandhinagar 382 428, India}%
\affiliation{Homi Bhabha National Institute, Anushaktinagar, Mumbai 400 094, India}%
\author{Sanket Patil,}%
\affiliation{Department of Physics, University of Wisconsin-Madison, WI 53706-1390}%
\author{Alexander Khrabrov, Igor Kaganovich}%
\affiliation{Princeton Plasma Physics Laboratory, Princeton University, Princeton, New Jersey 08543, USA}%

\date{\today}

\begin{abstract}
Recently, S Patil et al.\cite{Patil_013059_PRR_2022} have reported the existence of an enhanced operating regime when a
low-pressure (5 mTorr) capacitively coupled discharge (CCP) is driven by a very high radio-frequency (60 MHz) source in the 
presence of a weak external magnetic field applied parallel to its electrodes. Their Particle-in-Cell (PIC) simulations show, that a   
 significantly higher bulk plasma density and ion flux can be achieved at the electrode when the electron cyclotron 
frequency equals  half of the applied RF frequency for a given fixed voltage. In the present work we take a detailed look at this
phenomenon and further delineate  the effect of this ``electron bounce cyclotron resonance (EBCR)" on the electron and ion dynamics of the system.
We find that  the ionization collision rate and stochastic heating is maximum under resonance condition. 
The electron energy distribution function also indicates that the population of tail end electrons 
is highest for the case where EBCR is maximum. Formation of electric field transients in the bulk plasma region are also seen at lower values of applied 
magnetic field. Finally, we demonstrate that the EBCR induced effect is a low pressure phenomenon and  weakens as the neutral gas pressure increases. 
 The potential utility of this effect to advance the operational performance of CCP devices for industrial purposes is discussed.
\end{abstract}


\maketitle

\section{Introduction}
\label{intro}
\noindent The ion flux and the impact energy of ions incident on a substrate are two important parameters that control 
the processing rates and the quality of the wafer in plasma reactors used for silicon wafer etching and material 
deposition in semiconductor industries. The capacitively coupled plasma (CCP) reactor is the most popular tool 
for etching purposes in industries. It is typically operated by a 13.56 MHz radio-frequency (RF) source 
\cite{Lieberman_NJ_2005, Chen_APL_93_2008, Buddemeier_APL_67_1995}. The etching process is a function of the 
operating neutral gas pressure, the input power and the geometry of the system. Thus, both the ion flux and the ion 
energy changes simultaneously with input power for a fixed neutral gas pressure and system geometry. Accordingly, 
several detailed studies report that in the single frequency capacitively coupled plasma (SF-CCP) discharges the 
ion flux and the ion energy cannot be controlled independently \cite{Kaganovich_IEEE_34_2006, Kawamura_POP_13_2006, Lieberman_IEEE_16_1998, Godyak_SJP_78_1976, Godyak_JAP_57_1985, Kaganovich_PRL_89_2002, 
Sharma_POP_21_2014, Sharma_PSST_22_2013, Sharma_POP_20_2013}. To overcome this limitation, several 
alternate ideas have been proposed in the past. One widely used idea in semiconductor industry is the dual-frequency 
capacitively coupled plasma (DF-CCP) device in which the low frequency ($f_l$) component controls the sheath width 
(i.e. effectively ion energy) while the high frequency ($f_h$) component influences the plasma density (i.e. the ion flux) \cite{Goto_IEEE_6_1993, Robiche_JPDAP_36_2003, Kim_POP_10_2003, Turner_PRL_96_2006, Sharma_JPDAP_46_2013, Karkari_APL_93_2008, Boyle_JPDAP_37_2004, Sharma_JPDAP_47_2014, Sharma_Thesis_2013}. However, frequency coupling occurs 
if these two frequencies (i.e. $f_l$ and $f_h$) are too close to each other and as a result, independent control of ion flux 
and ion energy is not possible \cite{Kawamura_POP_13_2006, Turner_PRL_96_2006, Gans_APL_89_2006, Schulze_JPDAP_40_2007, Zhen_PSST_22_2013}. In order to minimize the frequency coupling effect, one has to 
choose $f_h$ very high compared to $f_l$, which gives rise to electromagnetic effects, another undesirable physical phenomenon.
 These electromagnetic effects (typically occurring at $f_h$ $>$ 70 MHz) create nonuniformity in the plasma, which ultimately 
 degrades the wafer quality \cite{Lieberman_PSST_11_2002, Perret_APL_83_2003, Perret_APL_86_2005}. Other 
 alternative schemes to achieve the goal of independent control of ion flux and ion energy are electrical asymmetric 
 effects \cite{Czarnetzki_PSST_20_2011, Heil_JPDAP_41_2008, Bruneau_PSST_23_2014, Bruneau_PRL_114_2015, Schungel_JPDAP_49_2016} and non-sinusoidal, tailored voltage/ current waveform excitations 
 \cite{Economou_JVST_31_2013, Lafleur_PSST_25_2016, Qin_PSST_19_2010, Shin_PSST_20_2011, 
 Sharma_PSST_24_2015, Sharma_PSST_29_114001_2020, Sharma_POP_28_103502_2021}. In addition to that, 
 recently researchers have also explored the consequences of driving a  CCP by frequencies up to a few tens of MHz or even higher. 
A CCP driven in the very high frequency (VHF) band (i.e. 30-300 MHz) creates higher plasma density because of the 
increase in the plasma current for a fixed discharge power and lower DC self-bias, thereby giving a higher etching rate with 
lower damage compared to the low frequency plasma excitation. Interesting phenomenon like formation of an electron beam near to 
the sheath edge which travels across the bulk plasma without collisions and interacts with the opposite sheath can 
create ionization and sustain the plasma in VHF driven CCP discharges \cite{Shahid_PSST_19_2010, Wilczek_PSST_24_2015, Sharma_POP_23_2016}. Other phenomena like the formation of electric field transients and its effect on the bulk 
plasma and sheaths, the presence of higher harmonics in voltage and current etc. are published in the literature of very high frequency CCP 
discharges \cite{Sharma_POP_23_2016, Sharma_POP_25_080705_2018, Sharma_JPDAP_52_2019, Miller_PSST_15_2006, Upsdhyay_JPDAP_46_2013, Sharma_POP_25_063501_2018, Wilczek_PSST_27_2018, Sharma_POP_26_2019, Sharma_PSST_29_2020}. 
 
 External magnetic fields have been used in many previous studies in the area of magnetically enhanced reactive ion 
 etching (MERIE) \cite{Muller_ME_10_1989, Lieberman_IEEE_19_1991, Hutchinson_IEEE_23_1995, Kushner_APL_94_2003, Vasenkov_JAP_95_2004, Park_IEEE_25_1997}. You et al. \cite{You_TSF_519_2011} experimentally investigated the 
 effect of static external magnetic field on asymmetric  single frequency CCP argon discharges. The operated frequency is 13.56 MHz 
 in low and intermediate pressure regime. They reported the appearance of E$\times$B drift resulting from the electric field E perpendicular to the
  electrodes and external magnetic field B parallel to the electrodes. In another PIC simulation study, conducted by 
  Yang et al. \cite{Yang_PPP_14_2017, Yang_PSST_27_2018}, an asymmetric magnetic field with variable gradients 
  has been used to create asymmetry in the arrangement of a capacitively coupled discharge reactor. They showed that the magnetic field asymmetry 
  offers a method to independently control the ion flux and the ion energy. Later, Sharma et al. 
  \cite{Sharma_POP_25_080704_2018} used a well-tested Electrostatic Direct Implicit Particle-In-Cell (EDIPIC) code \cite{Sydorenko_Thesis_2006, Campanell_POP_19_2012, Sheehan_PRL_111_2013, Carlsson_PSST_26_2017, Campanell_APL_103_2013, Charov_PSST_28_2019} and showed that enhancement in ion flux and effective control on ion energy can also be 
  obtained by applying a uniform transverse external magnetic field.

As discussed above, an external transverse magnetic field in SF-CCP enhances the ion flux and also provides a mechanism 
to control the ion flux and ion energy. However, the presence of a large magnetic field (typically few tens to hundreds of 
Gauss) introduces an undesirable non-uniformity in the bulk plasma \cite{Barnat_PSST_17_2008, Fan_POP_20_2013} and 
thus compromises the uniformity of the wafer. Barnat et al. \cite{Barnat_PSST_17_2008} studied the effect 
of an external magnetic field on 
argon CCP discharges at 13.56 MHz applied frequency, and observed that the transverse magnetic field causes 
non-uniformities in the discharge owing to the E$\times$B drift. Fan et al. \cite{Fan_POP_20_2013} used 2D-3V PIC simulations 
to investigate the effect of an externally applied transverse magnetic field on an argon CCP discharge at 20 mTorr 
pressure driven by a 13.56 MHz RF source with an applied voltage of 200 V. It was found that for uniform magnetic field, 
the plasma density increases by increasing magnetic field from 0 G to 50 G. But above 20 G, the non-uniformity in the 
plasma caused by the E$\times$B drift becomes significant.

SF-CCP driven by VHF in presence of external transverse weak uniform magnetic field enters in a new operational regime. 
Recently, a different regime where a significant enhancement in the performance of low-pressure CCP discharge at much 
lower transverse magnetic field has been reported \cite{Patil_arXiv, Patil_013059_PRR_2022, Zhang_PRE_104_045209_2021, Wang_PSST_30_10LT01_2021}.
 Patil el al. \cite{Patil_013059_PRR_2022, Patil_arXiv}  were the first to report, using PIC simulations,  that a significant enhancement in the 
performance of a low-pressure (5 mTorr) CCP discharge can be achieved by applying a very weak transverse magnetic
 field ($\sim$ 10 G) at the applied frequency of 60 MHz. It was shown  that the enhancement in 
 performance is due to a resonance effect which occurs when the electron cyclotron frequency ($f_{ce}=eB/m_e$, where e, 
 $m_e$  and B are the electronic charge, mass and external applied magnetic field respectively) matches one half of 
 the applied 
 RF frequency ($f_{rf}$). Subsequently, Zhang el al. \cite{Zhang_PRE_104_045209_2021} also observed a similar 
 type of effect using PIC simulations in the 
 frequency range of 13.56 MHz to 60 MHz and then verified these findings experimentally. Recently 
 Wang et al. \cite{Wang_PSST_30_10LT01_2021}
  have reported the attenuation of self-excitation of plasma series resonance (PSR) in 
 the presence of a transverse external magnetic field using 1D-3V PIC simulation.

In this study, we provide a more detailed study of the effect of a weak transverse magnetic field ($\sim$10 G) on an argon CCP discharge 
driven at a very high frequency (60 MHz), at a very low neutral gas pressure (5 mTorr). We have used the 1D-3V 
Electrostatic Direct Implicit Particle-In-Cell (EDIPIC) code. We have shown that high plasma density, and thereby high 
ion flux can be achieved at a very low magnetic field for which the non-uniformity in plasma would be negligible. The 
paper is structured as follows. In Sec. \ref{SimulationTechnique}, we describe the simulation technique and parameters including boundary 
conditions. The simulation results at different magnetic fields and a discussion about the same are described in Sec. \ref{RandD}. 
The concluding remarks are presented in Sec. IV.

\section{Simulation Technique and Parameters}
\label{SimulationTechnique}
\begin{figure}[htp]
\center
\includegraphics[width=8cm]{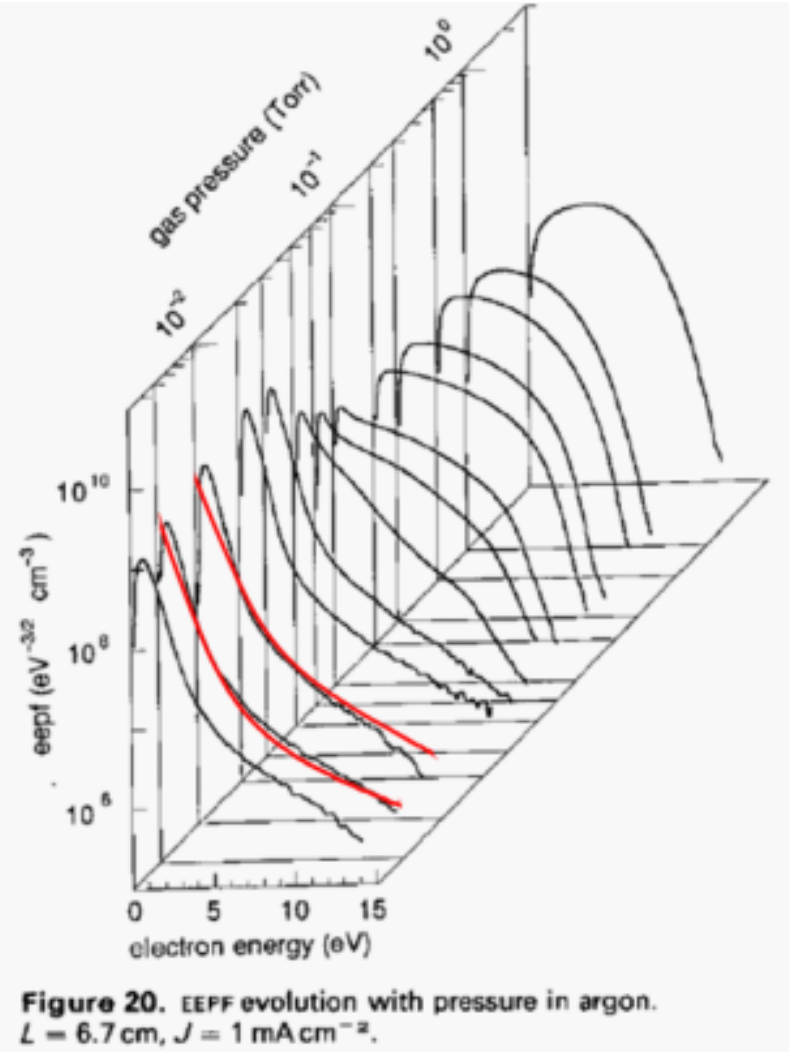}
\caption{Comparison of electron energy distribution function (EEDF) from experiments by Godyak et al.\cite{Godyak_1992} and simulation 
using EDIPIC code. The red curve generated at 5 mTorr and 10 mTorr by EDIPIC is in good agreement with experimental results.}
\label{fig:figure1}
\end{figure}

\noindent A well-tested and widely used 1D-3V Electrostatic Direct Implicit Particle-In-Cell (EDIPIC) code \cite{Sydorenko_Thesis_2006, Campanell_POP_19_2012, Sheehan_PRL_111_2013, Carlsson_PSST_26_2017, Campanell_APL_103_2013, Charov_PSST_28_2019, Sharma_POP_25_080704_2018} has been used for this study. 
It simulates the plasma domain bounded by two parallel plate electrodes. The code is based on the well-established 
Particle-in-Cell/Monte Carlo Collision (PIC-MCC) technique \cite{Birdsall_AH_1991, Hockney_AH_1988}. The 
electron-neutral collisions considered here are elastic, excitation and ionization. For  the ion-neutral collisions, 
elastic and charge exchange collisions are taken into account. It is important to note that the metastable reactions are not critical at low 
pressure and thus not included here. The cross-section data used for the collisions have been taken from well-tested 
sources \cite{Shahid_JAP_82_1988, Lauro_JPDAP_37_2004}. The code evolves the positions and velocities of electrons 
as well as singly ionized argon ions i.e. Ar$^+$. It is to be noted that although the code is 1D in position, it can accurately 
simulate the E$\times$B motion of charged particles because it has 3D in velocity space. The neutral gas dynamics is not 
evolved; the neutral gas distribution is straightforwardly uniform between the electrodes throughout the simulation. The neutral gas temperature
also kept constant at 300 K. We have also ignored secondary electron emission since at low pressure, it has a negligible 
effect on discharge properties \cite{Lieberman_NJ_2005}. The external circuit has also not been considered here because 
in a simulation, it is easier to assume a given potential at the electrodes and allow the current form to adjust accordingly. 
We have also conducted a few simulations to benchmark EDIPIC code against a well established experimental paper 
\cite{Godyak_1992} to strengthen our present simulation results. The experimental parameters used in 
Godyak el al. \cite{Godyak_1992} are: L = 6.7 cm, J = 1 $mAcm^{-2}$, $f_{rf}$ = 13.56 MHz at different gas pressure for argon 
discharge. We have used EDIPIC for the same set of discharge parameters and compared electron energy distribution 
function (EEDF) at 5 mTorr and 10 mTorr. Figure \ref{fig:figure1} shows the comparison of EEDF measured by experiments and 
results from simulation (red curve) at these two pressures. It is clearly seen  that bi-Maxwellian nature of EEDF 
obtained from simulation (red solid line in figure \ref{fig:figure1}) is in good agreement with the experimental results.

The neutral argon gas at 5 mTorr pressure is used in present research work. The frequency and applied voltage amplitude are 
60 MHz and 100 V respectively. The voltage having the following waveform is applied between the grounded electrode (GE) and the 
so-called powered electrode (PE), which are 32 mm apart:
\begin{equation}
V_{rf} = V_0 sin ( 2\pi f_{rf}t+\phi ) ,
\label{appliedVoltage}
\end{equation}
Here $V_0$, $f_{rf}$ and $\phi$ are the voltage amplitude, applied radio-frequency and phase respectively. The electrode 
dimensions are anticipated to be significantly larger than the inter-electrode gap. Thus, the 1D spatial assumption is valid.  
The external magnetic field (B) has been varied from 0 G to 107 G and is applied parallel to the electrodes . The initial electron and ion temperatures 
are 2 eV and 0.026 eV (300 K) respectively. The cell size ($\Delta$x) used is 1/8 of the Debye length 
($\lambda_{De}=\sqrt{{\varepsilon}_0 T_e/n_0 e}$ where $\varepsilon_0$ is the permittivity of free space and $T_e$ is the electron temperature in eV) for a density of $5\times10^{15}$ $m^{-3}$. The cell size is therefore $2.62829\times10^{-5}$ m, which is small 
enough to resolve the Debye length ($\lambda_{De}$). The number of cells turns out to be 1217 for the system length of 32 mm. The time 
step ($\Delta t$) is estimated as $\Delta x$ / (maximum expected velocity), where the maximum expected velocity is four times the thermal 
velocity. The time step is thus $7.83208\times10^{-12}$ s, which satisfies the stability criterion $\omega_{pe}$ $\Delta t$ $<$ 0.2, since 
$\omega_{pe}$ $\Delta t$ = 0.03 $(\omega_{pe}=\sqrt{n_{0}e^{2}/\varepsilon_0 m_e})$. This time step appropriately resolves the cyclotron motion at the maximum magnetic field of 107 G as 
$T_{ce}$ = 1/$f_{ce}$ = $3.33\times10^{-9}$ s $\sim$ 425.6 $\Delta t$. Here $f_{ce}=eB/2\pi m_e$, where B and $m_e$ is the external 
magnetic field and the electronic mass respectively, is the cyclotron frequency. The number of super particles per cell is initially 400 which 
gives total number of particles of the order of $\sim$ $5\times10^5$. The external magnetic field (B) which is implemented parallel to the 
electrodes is quantified by the ratio $r = 2f_{ce}/f_{rf}$ so that the resonance can occur at r = 1. Here the value of `r' is varied from 
0 to 10 by varying the magnitude of external magnetic field from 0 G to 107 G. We have assumed the perfect absorbing boundary conditions here, 
i.e. all charged particles are absorbed when they reach at the electrodes.

\section{Results and Discussion}
\label{RandD}

\subsection{Plasma density and ion flux}
\label{plasmaDensityIonFlux}
The higher etch rate can be obtained by getting higher ion flux at the electrode for which, plasma density has to be increased. 
One of the ways to achieve this is to use a higher neutral gas pressure in the CCP. However, operating the CCP at a low pressure 
(typically at a few mTorr) is an essential criterion to restore anisotropy 
in the etching process by creating a collisionless plasma \cite{Lieberman_NJ_2005}.

\begin{figure}[htp]
\center
\includegraphics[width=16.0cm,angle=0]{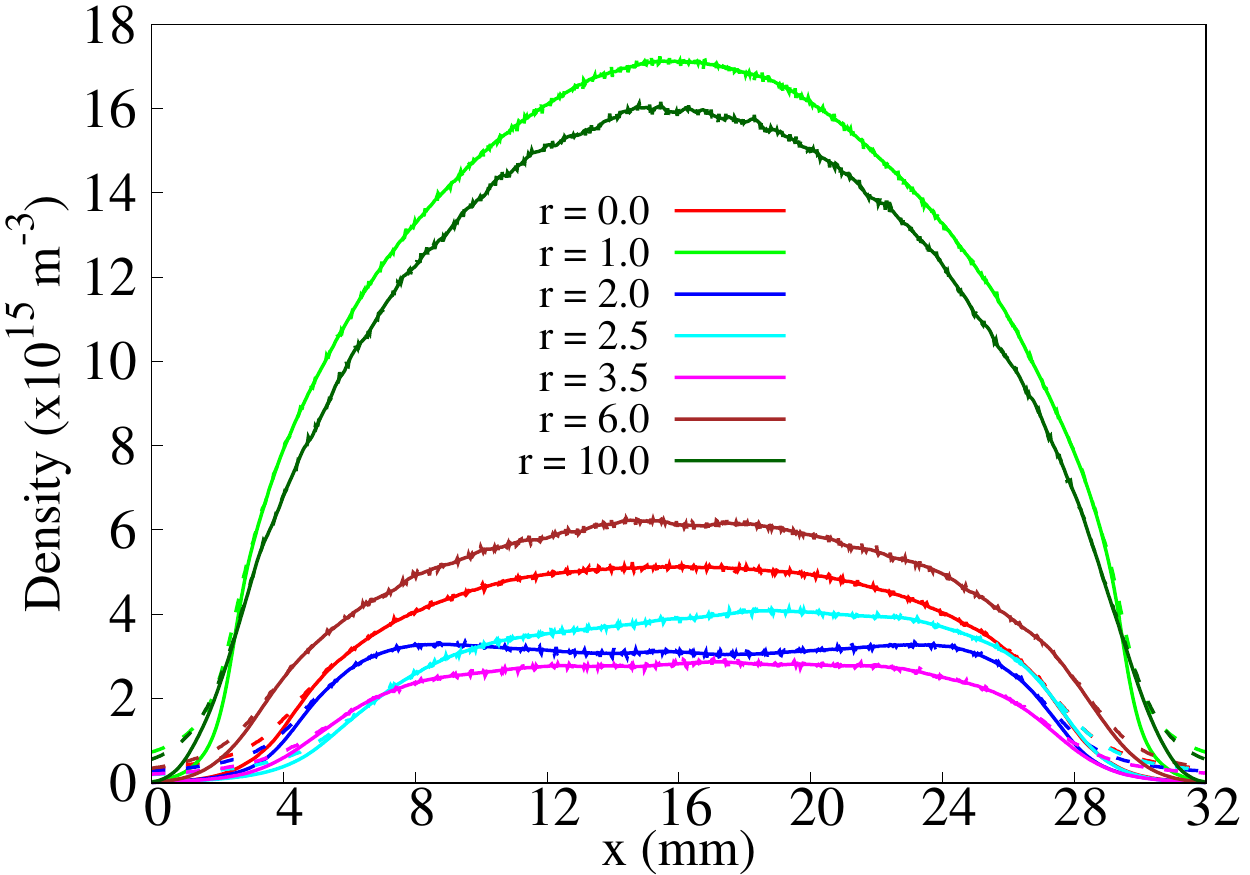}
\caption{Spatial variation of time average ion density ($n_i$, dashed line) and electron density ($n_e$, solid line) at different values of `r'.}
\label{fig:ElIonDen}
\end{figure}

The spatial variation of the electron and ion density averaged over last 100 RF cycles has been shown in figure \ref{fig:ElIonDen} for a few specific 
values of `r'. It is clear from the figure that the peak density is at r =1 (for B = 10.7 G) which is nearly 3.5 times higher than 
unmagnetized case i.e. r = 0 (red curve). The density decreases significantly for the higher values of r = 2, 2.5 and 3.5 compared 
to r = 1. Finally, for r $>$ 3.5, the density starts to increase with `r' and at r =10 (for B = 107.2 G) the density reaches nearly 
the value of the r =1 case. Also it is to be noted that the density profile is asymmetric at r = 2.5 case. We will discuss this phenomenon 
in a later section.

\begin{figure}[htp]
\center
\includegraphics[width=16cm]{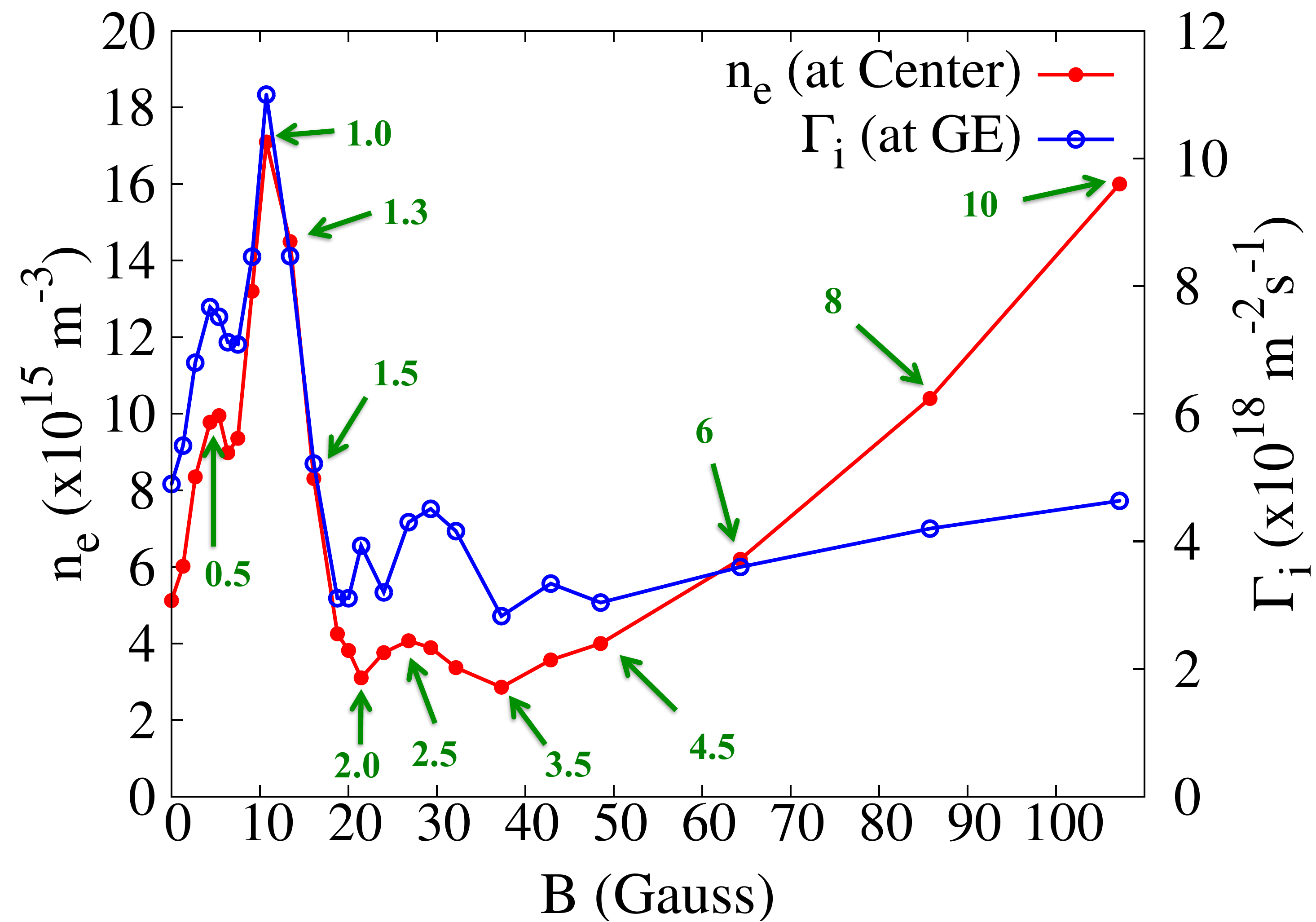}
\caption{Variation of peak electron density (left y-axis) in the discharge and ion flux (right y-axis) at the grounded electrode, with the 
external applied magnetic field.  The numbers shown (in green font) in the curve indicates the specific different values of `r' corresponding 
to external magnetic field.}
\label{fig:PeakElDensityIonFlux}
\end{figure}

Figure \ref{fig:PeakElDensityIonFlux} shows a plot of the peak electron density (at the center of discharge) and the ion flux 
(at the electrodes) against the magnetic 
field strength (i.e. against `r'). The left and right y-axis show the electron density and ion flux respectively. A few specific values of `r' are 
also shown in the figure by a green font. The results demonstrate that at particular values of `r' (i.e. 0.5 and 1), the plasma density and 
the ion flux display maximas. As shown in figure \ref{fig:ElIonDen}, at r = 1 a maximum plasma density of 
$1.7\times10^{16}$ $m^{-3}$ (at B = 10.7 G) 
appears and after that the density decreases rapidly up to r = 2. The density again increases monotonically after r = 3.5 (B = 37.3 G) and 
attains a value of $1.6\times10^{16}$ $m^{-3}$ at r =10 (B = 107.2 G). There is plenty of past literature that have reported the reason 
behind the monotonic increase of density beyond r = 3.5 in magnetically enhanced CCP devices \cite{Muller_ME_10_1989, 
Lieberman_IEEE_19_1991, Hutchinson_IEEE_23_1995, Kushner_APL_94_2003, Vasenkov_JAP_95_2004, Park_IEEE_25_1997}. 
In this paper the novel result at 
r =1 where a maximum in density is achieved at a very weak external magnetic field of $\sim$ 10.7 G is going to be discussed in detail. 

It is expected that the ion flux ($\Gamma_i = n_iu_i$) at the electrodes (or equivalently the ion current density $J_i = en_iu_i$) will 
follow a similar trend as the
density with a corresponding peak at r = 1, as is observed in figure \ref{fig:PeakElDensityIonFlux} (blue curve). However, the ion flux 
does not increase like the 
density above r = 3.5. This means that at a particular amplitude of the applied voltage, the ion flux obtained at r = 1 
($\sim$ $11\times10^{18}$ $m^{-2}s^{-1}$) is considerably higher compared to the case in the absence of magnetic field 
($\sim$ $5\times 10^{18}$ $m^{-2}s^{-1}$) as well as for r $>$ 3.5. So the larger magnetic fields (r $\sim$ 10) can create higher 
plasma densities although not proportionately higher ion flux at the electrodes.

\subsection{Ion energy and plasma potential}
\label{IonEnergyPlasmaPotential}
To find out whether the observed peak at r = 1 is an improvement in the overall performance of the discharge, the average energy of ions 
incident on the electrodes needs to be observed.

\begin{figure*}[htp]
\center
\includegraphics[width=16cm]{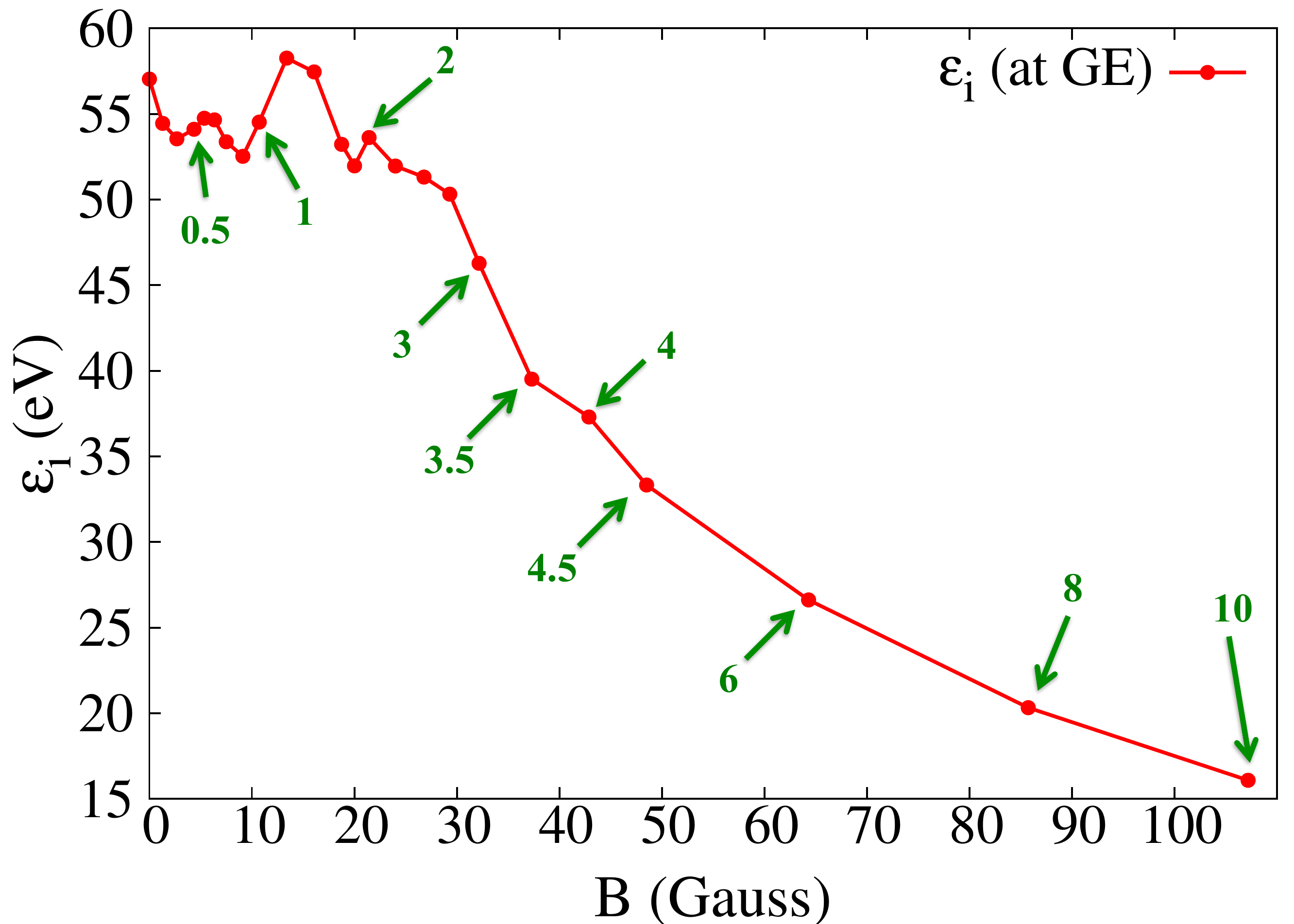}
\caption{Figure shows the variation of average energy of the incident ions on the grounded electrode with static external magnetic field. Few specific values of ‘r’ also shown here in green text.}
\label{fig:AvgIonEnergy}
\end{figure*}

Figure \ref{fig:AvgIonEnergy} demonstrates the change in the average energy of incident ions  on the grounded electrode (at 32 mm) with the 
strength of the external magnetic field. It can be seen that the average energy of ions at r = 1 (i.e. at 10.7 G) is 54.5 eV, which 
is less than 57.0 ev the value corresponding to the case when the magnetic field is absent. We have calculated the sheath width (also shown in figure \ref{fig:ElIonDen}) which is $\sim$5.2 mm and $\sim$3 mm for r = 0 and r =1 case respectively. Therefore the change in ion energy can be ascribed to the reduction of the sheath width that affects the potential drop across the sheath. We can conclude that at r = 1, a considerably greater ions flux with a somewhat lower average ion energy 
can be achieved at the electrodes compared to the case without an external magnetic field.
The energy of ions at r $\geq$ 3.0 is much lower than that at r = 1. This means that operation at r $\geq$ 3.0, has the advantage of lower ion
 energies, however at the cost of ion flux. Furthermore, the disadvantage of non-uniformity occurs in the plasma due to E$\times$B
  drift at higher magnetic fields, which would be negligible at r = 1. Consequently operating the CCP at r = 1 has a clear advantage 
  because of negligible non-uniformity in plasma.

\subsection{Rationale for choosing the magnetic field values}
\label{Rationale}
\begin{figure}[htp]
\center
\includegraphics[width=16cm]{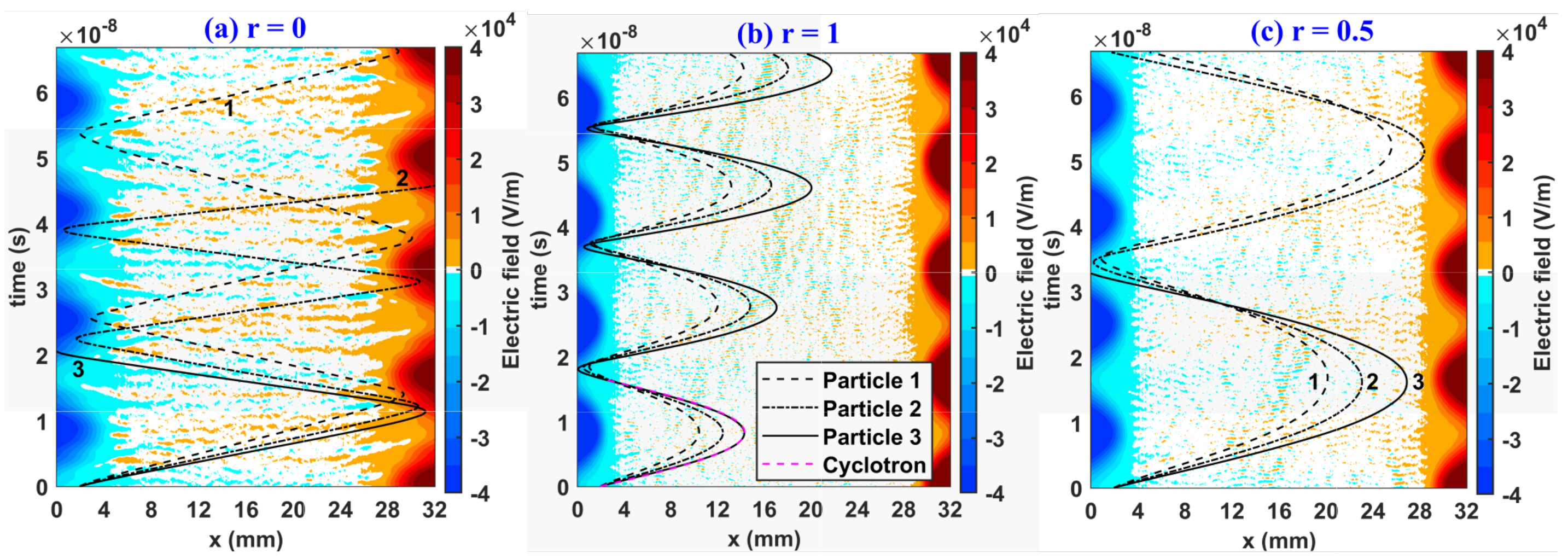}
\caption{Above figure reproduced from our earlier work by Patil et al. \cite{Patil_013059_PRR_2022}. Figure shows the trajectories of electrons, 
which are reflected from the expanding sheaths having 3 different energies. The trajectories 
are plotted onto the spatiotemporal profile of electric field. The initial energies of particles 1, 2 and 3 are 4 eV, 8 eV and 12 eV respectively. It is 
clear from figure that the three particles reach the opposite sheath at different times for r = 0, however for r = 0.5 and r = 1, the three particles return to the same sheath simultaneously. For r = 1, the dotted magenta line shows pure cyclotron motion (no RF fields) of a particle with energy 
12 eV. It is clear from figure that the trajectory of particle 3 closely resembles cyclotron motion.}
\label{fig:Trajectories}
\end{figure}

In this subsection, we examine the physical origin of this specific effect. In figure \ref{fig:Trajectories} we show trajectories of 3 electrons, which interact multiple 
times with the sheath.  This figure has been already reported in our previous publication \cite{Patil_013059_PRR_2022} 
and reproduced here for completeness. These 3 electrons are labelled as 1, 2 and 3 having initial energy of 4 eV, 8 eV and 12 eV 
respectively. The figure \ref{fig:Trajectories} (a), (b) and (c) are for case B = 0 G (i.e. r = 0), B = 10.7 G (corresponding to r = 1.0) 
and B = 5.3 G 
(corresponding to r = 0.5) respectively. It is clear from the figure that in each case, the electrons are bounced back into the 
bulk plasma by the strong oscillating electric field of the sheath when they move towards the electrode. The discharge here 
is nearly collisionless and the 
electron-neutral collision is not significant 
inside the bulk plasma. In figure \ref{fig:Trajectories} (a) for the r =0 case, particle 1 (i.e. 4 eV) reaches the opposite sheath 
after getting a kick from the expanding left sheath. 
It bounces back again to the left sheath because of getting a kick from the expanding right sheath and gets trapped between 
the two sheaths. It is eventually  lost to one of the
 electrodes after a few oscillations. In a similar way, particle 2 (i.e. the 8 eV one) is also trapped between the two sheaths 
 and eventually gets  lost to the right electrode 
 at 45 ns after a couple of oscillations. Finally, particle 3 (i.e. one with  12 eV) gets kicked by the expanding left sheath, reaches the right sheath, 
 bounces back and gets lost to the left electrode at $\sim$ 20 ns. The above results clearly show that electrons 
 having different energies or velocities 
 have different travel time between the two sheaths. Consequently, they are generally not in the resonant stage because their subsequent 
 interactions with the sheaths are not in the same phase as the sheath oscillation. In figure \ref{fig:Trajectories} (b) for r = 1, it is to be 
 noted that the electron motion closely resembles the pure cyclotron motion (the RF electric field inside bulk plasma is weak i.e. $<$ $10^3$ V/m). Consequently, all the 3 electrons getting kicked by the left sheath arrive to the same sheath after half a `cyclotron period' (shown by dashed 
 magenta line) irrespective of their velocities. It is because the `cyclotron period' is not a function of the velocity of the electron and there being no velocity component
 parallel to the electrodes at the time of impact with the sheath. The duration of the subsequent reflections or bounces is called the “bounce time” 
 here. For the case of “r=1”, the bounce time is equal to an RF period.  As a result, the electrons, which are kicked by an expanding phase of an 
 RF sheath, return to the same sheath at the same phase and undergo another energy gain from the interaction with expanding sheath. This mechanism for r = 1 expedites the effective production of energetic electrons as compared to r = 0 case. Consequently the electrons can gain substantial energy by this mechanism and contribute significantly to the process of ionization near to the sheath edge which then  results in an increase 
 of the ion flux. The electrons can remain in such a resonance till they collide with a neutral atom and that may occur once in every few RF 
 periods on an average at such low pressures. The generation of high energetic electrons can also be seen in the profile of the electron energy 
 distribution function (EEDF) and in the density of electrons having energy greater than the ionization potential. We will discuss this in the next subsection.

It is to be noted that similar, but  smaller boosts in electron density and ion flux are spotted at r = 0.5 (see Fig. \ref{fig:PeakElDensityIonFlux}). 
The resonance 
phenomenon is less efficient for this case for two reasons. Firstly, here the electrons experience collisions inside the bulk 
plasma because the electrons that get kicked from the expanding sheath return to the same sheath after two RF periods (see Fig \ref{fig:Trajectories} (c)). 
The mechanism here is same as for r = 1 case however the electrons traverse longer paths compared to the r = 1 case. Secondly, the 
Larmor radius $r_L=m_e v_{\perp}/eB$, where $v_{\perp}$ is the component of velocity perpendicular to B, is big enough to touch the opposite sheath at such a low magnetic field. The interaction of electrons with the opposite sheath can disturb the electron's trajectory and interrupt the resonance.

As will become apparent after going over the simulation results, the discharge properties do not vary monotonously as the magnetic field is varied.
 The variation is particularly interesting at magnetic fields lower than 30 G. It has been observed that discharge properties of interest such as the
  density and the ion flux at the electrodes have extrema at simple rational values of the ratio of the cyclotron frequency ($f_{ce}$) to the applied RF
   frequency ($f_{rf}$), especially at the lower values (viz. 0.5, 1). Thus, the magnetic fields used in the simulations have been varied by changing 
   the ratio $2f_{ce} / f_{rf}$. The reason for using this ratio is as follows: At typical electron velocities (a few times the thermal velocity at 
   $T_e = 2$ eV), the magnetic force on the electron will dominate the electrostatic force because the electric field in the bulk is weak. The motion 
   of an electron thus closely resembles cyclotron gyration. An electron reflected from a sheath at some phase of an RF cycle will undergo a motion as shown in Fig. \ref{fig:Trajectories}. If the electron were to complete half a cyclotron gyration in one RF period, it would be incident on the sheath at
    the same phase as its previous incidence (Fig. \ref{fig:Trajectories} (b)). Such synchronization can go on either until the 
    electron undergoes collisions
     within the bulk or until it gains enough energy to reach the opposite sheath because of its large Larmor radius. This, and similar possible synchronization conditions (as shown in Fig. \ref{fig:Trajectories}) can be quantified by the number of half-cyclotron gyrations the electron 
     undergoes in one RF cycle. In terms of time periods, this quantity (henceforth denoted by ‘r’) should satisfy:  
\begin{equation}
r\frac{T_{ce}}{2} = T_{rf}
\label{eq2}
\end{equation}  
\begin{equation}
r = \frac{2T_{rf}}{T_{ce}}=2\frac{f_{ce}}{f_{rf}}
\label{eq3}
\end{equation}
Since $f_{ce}$  (=$eB/m_e$ ) is directly proportional to B, `r' is also directly proportional to B. The values of the magnetic field for all simulations
 have been determined by setting the value of `r' to the simple ratios (0.5, 1, 2, etc.) that are of interest. 

\subsection{Ionization rate}
\label{IonizationRate}

\begin{figure*}[htp]
\center
\includegraphics[width=16cm]{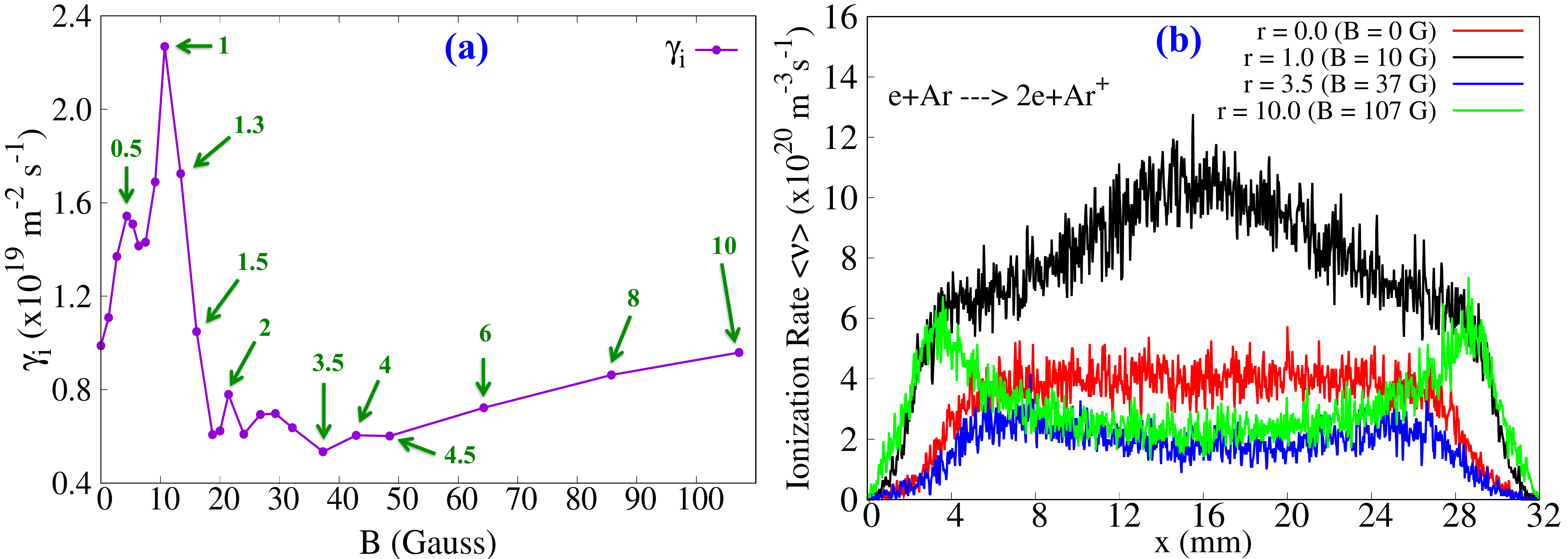}
\caption{Shows the variation of the (a) total ionization rate ($\gamma_i$) in the simulation regime with magnetic field at different values of `r' i.e. 
external magnetic field B and (b) ionization rate with respect to system length for specific values of `r' viz. 0, 1, 3.5 and 10.}
\label{fig:figure6}
\end{figure*}

Discharge properties are determined by particle and energy balance. To explain the variation of the properties hitherto discussed, these 
balances ought to be analyzed. The particle balance depends on the outgoing ion and electron flux and the ionization in the discharge. 
The variation of total ionization rate (ionization events per second per unit electrode area) in the simulation region between two electrodes 
versus strength of external magnetic field is shown in figures \ref{fig:figure6} (a). The different values of `r' (in green font) can also be observed in this figure. It is clearly observed that the total ionization rate is maximum for r = 1 where the plasma density is highest for the chosen 
parameters in our present study (see Fig \ref{fig:PeakElDensityIonFlux}). This is the case where highly energetic electrons are maximum in the system, which we will see in the description of electron energy distribution function (EEDF) in the next subsection. The total ionization rate decreases 
drastically above r = 1 and it again shows increasing trend above ‘r = 3.5’. However even at higher values of `r' the total ionization 
rate in the discharge region is much less compared to the ‘r = 1’ case. Figure \ref{fig:figure6} (b) demonstrates the profile of ionization rate with 
respect to system length at different values of `r' i.e. 0, 1, 3.5 and 10. It is clear from this figure that the ionization rate is maximum 
for r = 1 case where the electrons perform bounce cyclotron resonance phenomena as shown in figure \ref{fig:Trajectories} (b). However for the 
unmagnetized case i.e r = 0, the ionization  occurs uniformly in the bulk regime. For higher values of `r' (i.e 10) the electrons 
reflected from the oscillating sheath are tightly bound with the magnetic field lines and not able to penetrate to the center of discharge 
and hence create ionization in the vicinity of the sheath region only.

\subsection{Electron Energy Distribution Function}
\label{ElectronEnergyFunction}
The plasma parameters like electron density and temperature can be easily estimated by measuring the Electron Energy Distribution Function 
(EEDF), which is the most critical factor in chemical reactions during the plasma processing. Controlling the EEDF is very crucial to advance 
and optimize the etching process in CCP devices. There are many studies which show the change in EEDF with changing 
operating conditions in both magnetized \citep{Fan_POP_20_2013, You_TSF_519_2011, Hutchinson_IEEE_23_1995} and unmagnetized CCP discharges \citep{Fattah_APL_83_2003, Fattah_POP_19_2012, Fattah_POP_20_2013, Sharma_POP_23_2016, Sharma_POP_26_2019}. 
In this section, we investigate the effect of the variation of  the strength of external magnetic field on the shape of EEDF.

It is worthwhile to analyze the dependence of energy distribution of electrons on magnetic field in order to explain the variation in the
ionization rate.
\begin{figure}[htp]
\center
\includegraphics[width=8.5cm]{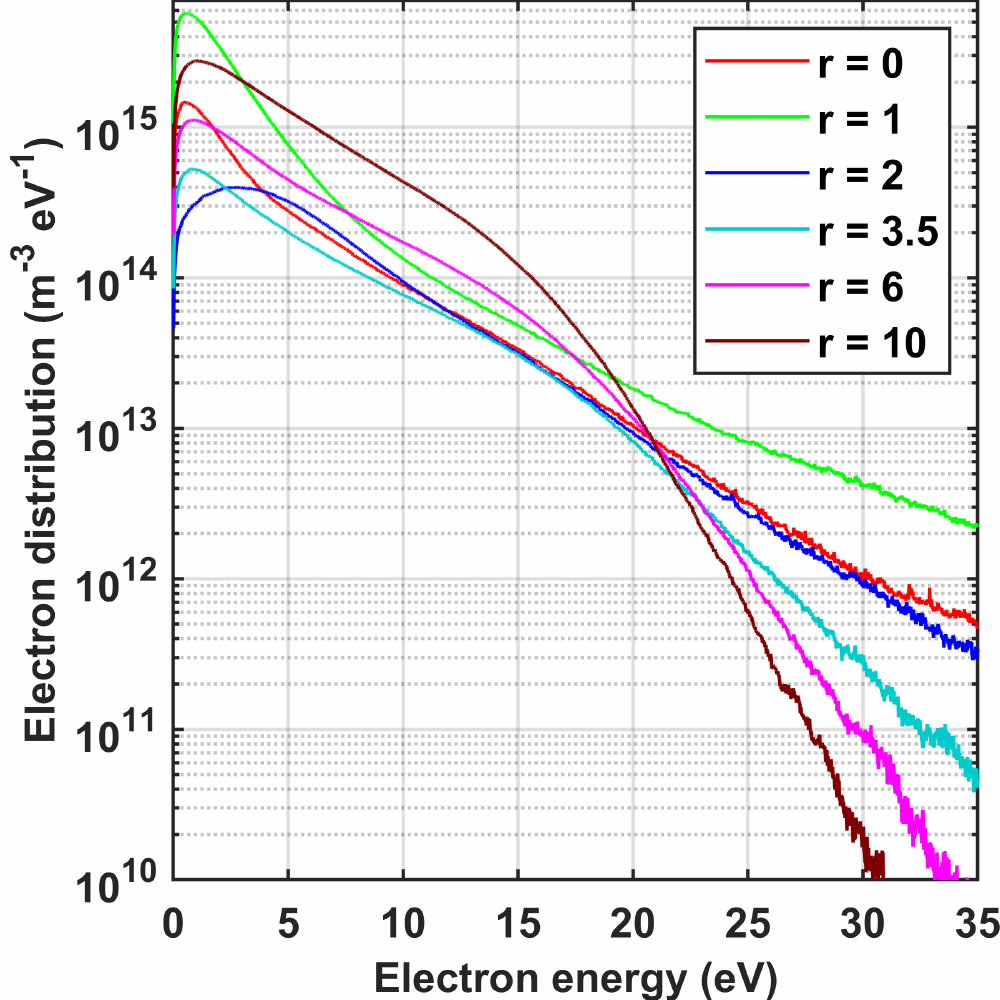}
\caption{Above figure demonstrates the variation of the shape of Electron energy distribution function (EEDF) at the centre of discharge for different values of `r'.}
\label{fig:figure7}
\end{figure}
Figure \ref{fig:figure7} shows the electron energy distribution function at the center of discharge for few specific different values of `r'. 
For r = 0 (i.e. unmagnetized case), the EEDF is weakly bi-Maxwellian in nature (the two slopes of curve intersects at $\sim$3.5 eV) which 
indicates the presence of two distinct electron populations with different temperatures. The bi-Maxwellian nature of the EEDF is owing to 
the collisionless or stochastic electron heating which mainly occurs in the vicinity of the sheath edges 
\cite{Godyak_PRL_65_996_1990, Kaganovich_IEEE_20_66_1992, Schweigert_PRL_92_155001_2004}. 
For the unmagnetized case, the density at the center of discharge is $5.1\times 10^{15}$ $m^{-3}$. A significant population ($\sim$ $3.3\times10^{15}$ $m^{-3}$ $\equiv$ $65 \%$) is of low energy electrons (bulk electrons), i.e. up to 3.5 eV. The population of 
medium energy electrons (between 3.5 eV and 16 eV i.e. the ionization potential) is $\sim$ $1.7\times10^{15}$ $m^{-3}$ $\equiv$ $33 \%$, 
and the remaining population ($\sim$ $1.1\times10^{14}$ $m^{-3}$ $\equiv$ $2 \%$) is of electrons in the high energy tail of the EEDF. 
As the value of `r' increases from 0 to 1, the shape of EEDF remains bi-Maxwellian, but the electron density in the bulk increases rapidly; 
for r = 1, the density at the center of discharge is $1.7 \times 10^{16}$ $m^{-3}$. As a result, the population of both low energy electrons 
($\sim$ $1.3 \times 10^{16}$ $m^{-3}$ $\equiv$ $77\%$) and high energy electrons in the tail increases 
($\sim$ $2.2 \times 10^{14}$ $m^{-3}$ $\equiv$ $1.3\%$). This is consistent with the increase in ionization rate.

For r = 2 the bounce cyclotron resonance phenomenon is not effective and hence the population of tail end electrons decreases and 
hence the overall density goes down which is clearly reflected in the EEDF curve which is Maxwellian. Figure \ref{fig:figure7} shows that for r = 3.5, 6 
and 10 the population of high-energy tail electrons depletes because the energetic electrons reflected from sheath are tightly bound 
with the magnetic field lines and cannot penetrate up to the center of discharge and hence most of the ionization occurs in the vicinity 
of the sheath edge as discussed in the previous section. The shape of EEDF is more like a Druyvesteyn for r = 6 and 10.

\begin{figure}[htp]
\center
\includegraphics[width=16cm,height=20.0cm]{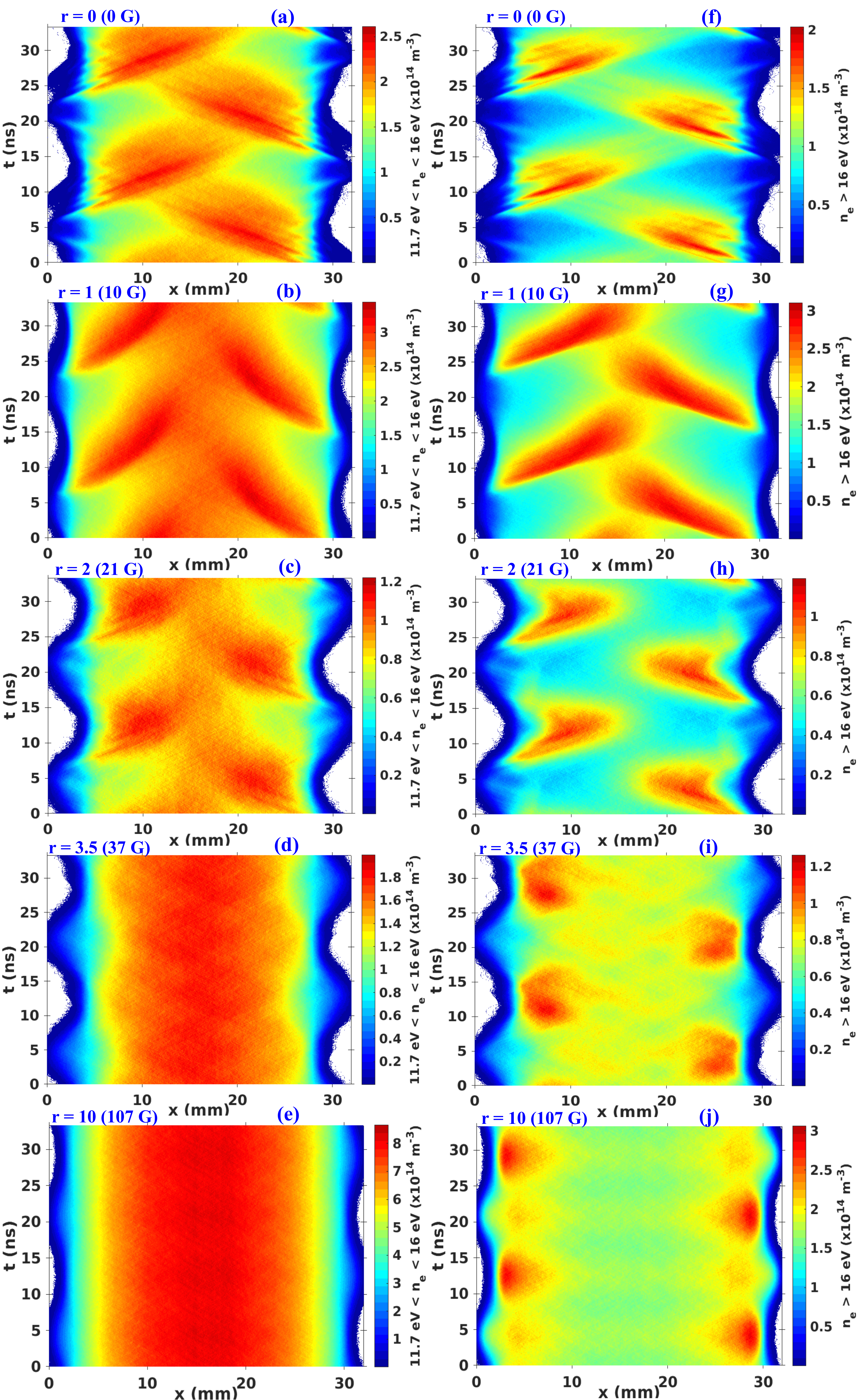}
\caption{Above figure shows the density of electrons having energy between 11.7 eV and 16 eV from (a)-(e) and  energy greater than 
16 eV from (f)-(j) for different values of `r' and corresponding magnetic field as indicated in the subplots.}
\label{fig:figure8}
\end{figure}

\begin{figure}[htp]
\center
\includegraphics[width=16cm]{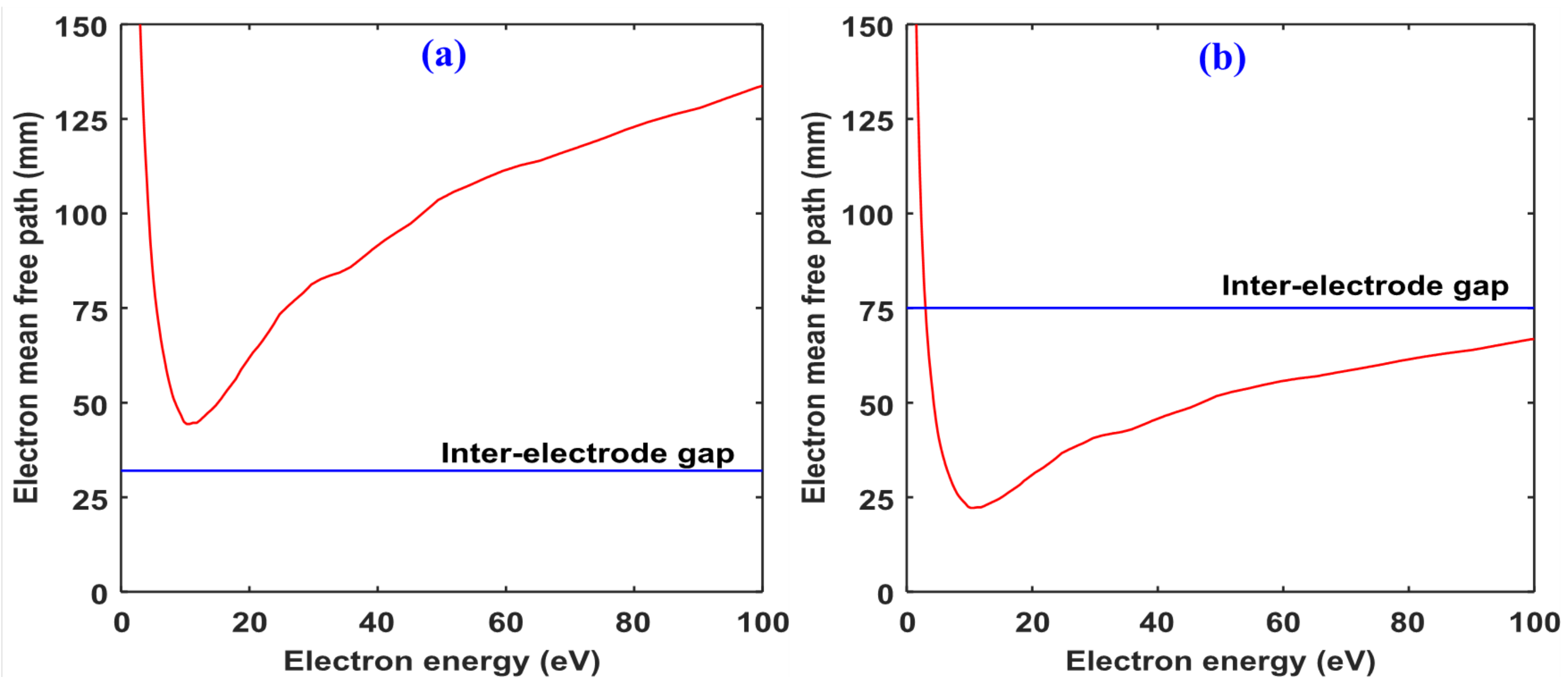}
\caption{(a) Electron mean free path in the current study, (b) Electron mean free path in Turner et al. \cite{Turner_PRL_76_2069_1996}.}
\label{fig:figure9}
\end{figure}

The change in the density of high-energy electrons by changing the value of `r' can be explained from the spatio-temporal profile of energetic 
electrons for two RF cycles in figure \ref{fig:figure8}. Sub figures (a)-(e) and (f)-(j) shows the density of electrons having energy between 
11.7 eV to 16 eV and greater than 16 eV respectively. Here the 11.7 eV is the excited state and 16 eV is the ionization potential of argon. 
It is clear from sub-figure (a) and (f) that for r = 0 (0 G), the density of electrons having energy between 11.7-16 eV and greater than 
16 eV are $\sim$ $2.5\times10^{14}$ $m^{-3}$ and $\sim$ $2.0\times10^{14}$ $m^{-3}$ respectively. It indicates that the 
energetic electrons emerge from the sheath edge at the time of sheath expansion and penetrate to the center of the discharge inside 
the bulk plasma. When the magnetic field increases to 10 G (r = 1) the density of such electrons increases significantly to  
$\sim$ $3.3\times10^{14}$ $m^{-3}$ (see sub-figure (b)) and $\sim$ $3.1\times10^{14}$ $m^{-3}$ (see sub-figure (g)) which is 
32 $\%$ and 55 $\%$ increase respectively. The energetic electrons emerging from the expanding sheath edge are more intense and 
reach at the center of discharge. This is a clear evidence that the bounce cyclotron-resonance phenomenon increases the density of tail 
end electrons (energy greater than ionization potential), which enhance the ionization process mainly at center (see figure \ref{fig:figure6} (b)) 
and results in maximum density formation in this case. When the value of r is increased to r = 2 (21 G), the electron density drastically 
decreases to $\sim$ $1.2\times10^{14}$ $m^{-3}$ (for energy range 11.7-16 eV) and $\sim$ $1.18\times10^{14}$ $m^{-3}$ 
(for $>$ 16 eV) respectively. It is nearly 63 $\%$ and $\sim$ 62 $\%$ decrease for 11.7-16 eV and greater than 16 eV respectively. 
Hence the overall density for this case decreases significantly. The shape of the energetic electrons emerging from the expanding 
sheath also changes because of the magnetic field strength. Their penetration to bulk decreases because of magnetic field strength. 
When the value of `r' is further increased to 3.5 (i.e at 37 G) the energetic electron density between 11.7-16 eV and greater than 
16 eV are $\sim$ $1.98\times10^{14}$ $m^{-3}$ (see sub-figure (d)) and $\sim$ $1.27\times 10^{14}$ $m^{-3}$ 
(see sub-figure (i)) respectively. It is nearly 65 $\%$ and 6 $\%$ increase for 11.7-16 eV and greater than 16 eV respectively 
compared to previous case. Sub-figure (d) indicates that majority of the density of electrons having energy between 11.7 eV to 
16 eV are at the center of the discharge, however the peak density of electrons having energy greater than 16 eV are confined 
near to the sheath edge. Finally, when value of `r' is 10 (i.e. at 107 G) the energetic electron density between 11.7-16 eV and greater 
than 16 eV are $\sim$ $8.5\times 10^{14}$ $m^{-3}$ (see sub-figure (e)) and $\sim$ $3.04\times 10^{14}$ $m^{-3}$ (see sub-figure (j)) respectively. Again the increase in density is $\sim$ 330 $\%$ and $\sim$140$\%$ for 11.7-16 eV and greater than 16 eV 
respectively compared to previous case. This is also reflected in the curve of EEDF in figure \ref{fig:figure7}. Sub-figure (e) indicates 
that majority of the density of electrons having energy between 11.7 eV to 16 eV are at the center of the discharge, however the 
peak density of electrons having energy greater than 16 eV are confined near to the sheath edge only.
The increase in density as `r' increases from 0 to 1 can be explained as follows:
For r = 0 as well as for r = 1, the electrons moving toward an electrode are pushed back or “reflected” into the bulk plasma due to the presence of 
strong electric field in the sheath region. Let us refer to this sheath as ‘sheath 1’. Subsequently, such electrons will travel through the
 bulk and their motion will be influenced by the bulk electric field. This field is weak and can significantly affect only the trajectory of 
  the very low energy electrons. The electrons under consideration, i.e. those, which interact with a sheath, are likely to have 
 higher energies. Thus, the trajectories of a majority of the electrons under consideration are unaffected by the bulk electric field. 
 An electron travelling through the bulk may either do so without colliding with a neutral or it may undergo a collision, the nature 
 of which (elastic, excitation or ionization) will be determined probabilistically by the energy of the electron. The former possibility 
 is more likely since the mean free path for collisions is larger than the inter-electrode gap (see figure \ref{fig:figure9} (a)).

If the electron does not collide at all in the unmagnetized case, it will reach the opposite sheath (see figure \ref{fig:Trajectories} (a)). Let us refer 
to it as ‘sheath 2’. The relative phase at which this electron arrives at sheath 2 with respect to the phase at which it was reflected 
by sheath 1 is determined by the time electron takes to travel the distance between the two sheaths. This travel time is thus 
estimated by the velocity of the electron after reflection and the strength of the electric field it encounters in the bulk plasma, but the later 
does not affect the trajectory significantly so has a 
negligible effect. As the electrons reflected by sheath 1 are distributed over 
the entire velocity range, their travel times must also be distributed over a certain range. For example, an electron having a higher 
velocity after reflection will reach sheath 2 earlier. This is true for electrons reflected from sheath 1 at all phases of an RF cycle. 
As a result, at each phase, reflected electrons in a certain velocity range will reach sheath 2 at a phase when the sheath width is 
minimum. These electrons constitute a fraction of the electrons absorbed by the electrode near sheath 2. The remaining fraction 
of the electrons that are absorbed consists of two types of electrons: one, electrons reflected by sheath 1 whose trajectories are 
changed due to collisions with neutrals such that they arrive at sheath 2 when the sheath width is minimum and two, the electrons 
created through ionization which also arrive at sheath 2 at this phase.
Now consider the trajectory of the reflected electrons in the r = 1 case as shown in figure \ref{fig:Trajectories} (b). If an electron does 
not collide with a neutral, its motion is affected by the magnetic field as well as the bulk electric field. However, the effect of the later 
is negligible (as was the case in the absence of magnetic field) and the electron motion can be approximated as cyclotron motion. 
Thus, the electron will return to sheath 1 after finishing half a gyration. However as explained earlier, the condition r = 1 means that 
the duration of half a gyration is accurately equal to the duration of an RF cycle. As the cyclotron frequency ($f_{ce}$) is independent  
of velocity, an electron reflected by sheath 1 will return to sheath 1 at the same phase as its reflection, regardless of its velocity 
after reflection. Consequently, a fraction of all the electrons reflected by a sheath at any given phase of an RF cycle return to the same sheath 
at the same phase and the remaining fraction collides with neutrals in the bulk. Consequently, the only electrons that will get absorbed 
by an electrode at the minimum-width phase of the corresponding sheath are either the ones undergoing collisions in the bulk or the ones 
created through ionization, both of which must arrive at the sheath edge when the sheath width is minimum.

Note that for the synchronization between electron motion and the RF cycle to be effective, mean free path of the electrons ($\lambda_e$) must 
be larger than the inter-electrode gap (i.e. collisionless discharge). If a majority of the reflected electrons collide with neutrals in the bulk, 
high density as achieved in the current study cannot be observed. For example, Turner et. al. \cite{Turner_PRL_76_2069_1996} have used an inter-electrode gap of 75 mm with a neutral gas pressure of 10 mTorr, for which the mean free path of a majority of electrons is smaller than the
 inter-electrode gap (see figure \ref{fig:figure9} (b)). Thus, they have not observed a high-density discharge at r = 1. Now if such a 
 low-frequency discharge were to be operated in the collisionless regime (narrower gap and lower pressure), there is another 
 reason why the synchronization would be ineffective. 
 
We have assumed that an electron reflected by sheath 1 will not interact with sheath 2 during its half gyration, which it can if its 
Larmor radius is large enough. For r = 1, the gap between the two sheaths in the present study is about 26 mm 
(see figure \ref{fig:Trajectories}) and the cyclotron frequency is 30 MHz. Therefore, an electron having Larmor radius greater than 26 mm 
must have energy greater than 68 eV. However the population of electrons in this energy range is realistically non-existent and can be ignored. 
This is what makes the synchronization between electron motion and the RF cycle effective. However, this synchronization would 
not lead to an increase in density if a substantial fraction of electrons 
reflected by sheath 1 interacted with sheath 2. Continuing the earlier 
example of Turner et al. \cite{Turner_PRL_76_2069_1996}, consider a low frequency discharge (i.e. 13.56 MHz) being operated with a narrow 
inter-electrode gap, about 32 mm. In that case, at r = 1, $f_{ce}$ = $f_{rf}/2$ = 6.78 MHz. Considering 3 mm sheath width, the Larmor 
radius for a reflected electron would have to be 26 mm to reach the other sheath, which corresponds to an energy of 3.5 eV. Consequently, 
after reflection a substantial fraction of electrons would reach the other sheath, which will render the synchronization ineffective. 
In conclusion, the synchronization is only effective for low-pressure, high frequency collisionless (or nearly collisionless) plasma.

The increase in density as `r' increases above r = 4 is simply because of the increasing confinement leading to slower transport of 
electrons i.e. lesser loss. Above a certain `r' (r $\sim$ 4), the Larmor radii for even the most energetic electrons 
having a non-negligible population are comparable to the sheath width (for example, at r = 4, $r_L$ $\sim$ 4.7 mm for a 35 eV electron). 
As a result, in the half cycle during which the width of a sheath decreases, a significant fraction of the reflected electrons will 
complete an entire gyration without approaching the electrode when the width of the corresponding sheath is minimum. 
Thus, at r $\geq$ 4, the electrons that do approach the electrode and get absorbed will mostly consist of the electrons 
undergoing collisions with neutrals. Consequently, the loss is determined by electron transport parameters. Without a magnetic field, 
the time and space scales of electron-neutral collisions (and thus of electron transport) are given by the collision frequency 
$f_{coll} = n_g \sigma v_e$ and the mean free path $\lambda_e = 1/(n_g \sigma)$. 
Here, $\sigma$ is the combined cross section 
for all types of electron-neutral collisions. The effective velocity $v_{eff} = \lambda_e f_{coll}$ is thus simply equal to the electron 
velocity $v_e$. However, when the magnetic field is such that the cyclotron frequency $f_{ce}$ is significantly greater than 
the collision frequency, the spatial scale of the transport is changed to twice the Larmor radius $r_L$ of the electron because 
until it collides, it will stay within a distance $2r_L$. The effective velocity now becomes 
$v_{eff}$ = $2r_L$ $f_{coll}$ = $(2v_e)/\omega_{ce}$ $f_{coll}$ = $v_e$  $f_{coll}/(\pi f_{ce})$. At r = 4, 
$f_{ce}$ = $1.2 \times 10^8$ Hz, which is much greater than even the maximum value of the collision frequency, 
$(f_{coll})_{max}$ $\sim$ $4.6 \times10^7$ Hz. Thus, the electron transport and thereby electron loss is significantly reduced by 
the magnetic field for r $>$ 4. This is why the density increases with `r'. The nature of the energy distribution of electrons 
(figure \ref{fig:figure6} a), which is different than that for r $<$ 2, may be attributed to the change in the heating mode: 
for r $>$ 3.5, the heating is dominated by ohmic heating due to an appearance of significant electric field in the bulk
(discussed in next subsection).

Between r = 2 and r = 4, the change in discharge properties with `r' is not trivial. There is clearly a transition in the 
qualitative nature of the heating and transport mechanisms. In this range of `r', a combination of the effects described 
above will determine the discharge properties. Other effects such as the change in sheath width and the corresponding 
change in collisionless heating may also play a significant role. Moreover, there is an asymmetry in the discharge properties 
in this range of `r', for example in density (see figure \ref{fig:PeakElDensityIonFlux}), which adds another degree of complexity. 
This will be explored fully in a separate article.

\subsection{Electron Heating}
\label{Heating}

There are two important mechanisms through which electrons gain energy in CCP discharges: ohmic (collisional) heating and stochastic 
(collisionless) heating \citep{Lieberman_NJ_2005}. Ohmic heating mainly occurs in the bulk plasma because of electron-neutral 
collisions and stochastic heating is localized near the sheath edges due to momentum transfer between electrons and the 
oscillating high voltage sheaths \cite{Kawamura_POP_13_2006, Kaganovich_IEEE_34_2006, Kaganovich_PRL_89_2002, 
Sharma_PSST_22_2013, Lieberman_IEEE_16_638_1988, Kawamura_POP_21_123505_2014, Turner_PRL_75_1312_1995}. 
There is a high potential difference across sheath in CCP discharges which is large compared to the energy of plasma electrons. 
As a result, the electrons are reflected and confined within the discharge. The effect of external magnetic field on heating in an 
RF-CCP discharges was reported by Turner et al. \cite{Turner_PRL_76_2069_1996}. Using a fluid approach, they showed that 
collisionless heating could be allied with the rarefaction and compression of electrons flowing through the inhomogeneous plasma 
medium, which is stated as the ‘pressure heating’ mechanism. They reported the existence of a heating mode transition from 
collisionless to ohmic in the presence of weak external transverse magnetic field.

\begin{figure}[htp]
\center
\includegraphics[width=16cm]{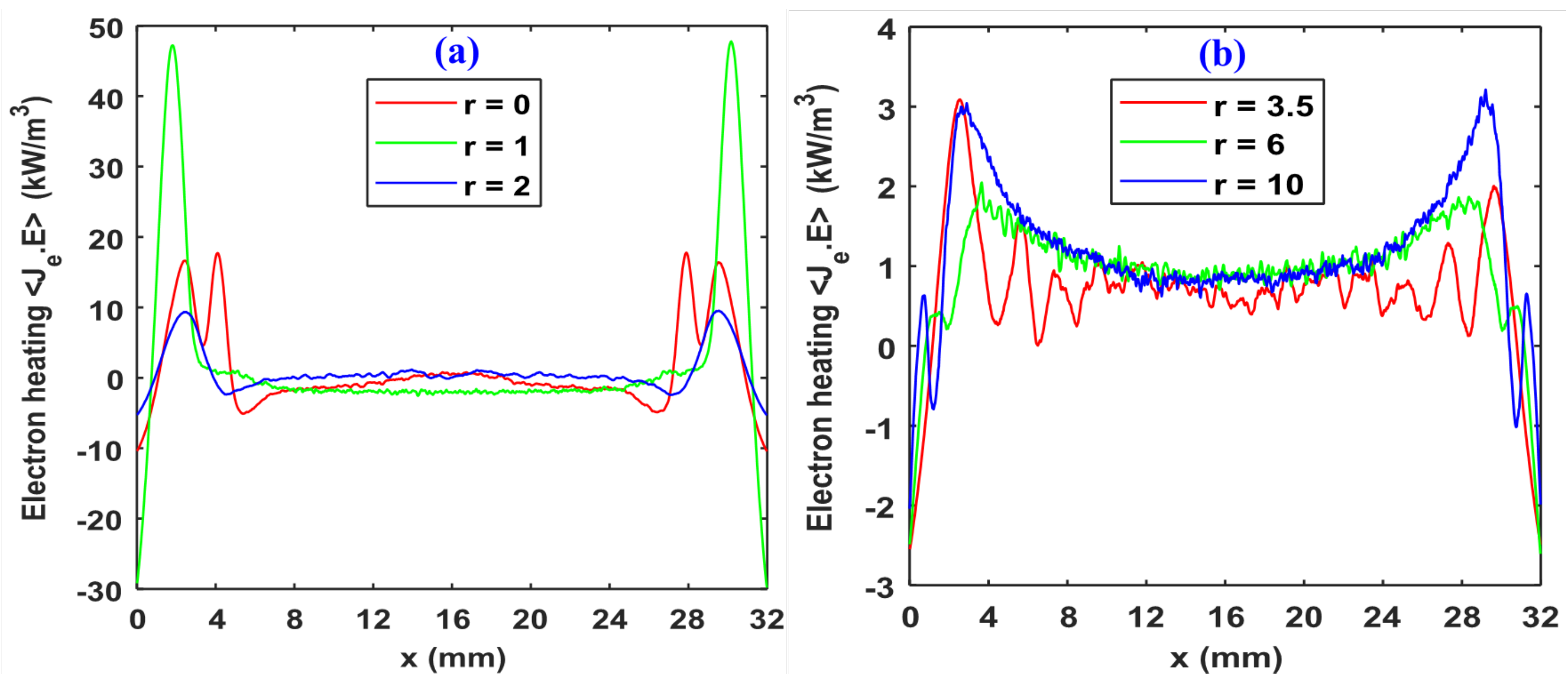}
\caption{Time average electron heating for some specific values of ‘r’ i.e. (a) r = 0, 1, 2 and (b) r = 3.5, 6 and 10.}
\label{fig:figure10}
\end{figure}

Figure \ref{fig:figure10} shows the variation of time-averaged electron heating in the discharge for some specific values of ‘r’.
It is clear from figure \ref{fig:figure10} (a) that the positive heating is localized in the sheath regions and a small positive or negative 
electron heating is present in the bulk plasma. The physical explanation of the presence of negative electron heating 
i.e. $<J.E>$ can be found in literature \cite{Kaganovich_IEEE_34_2006, Surendra_PRL_66_1479_1991, Turner_JPDAP_42_194008_2009}.
For r = 1, the heating is maximum and is much higher compared to unmagnetized case. Here the dominant heating mechanism is 
stochastic or collisionless. This effect diminishes or disappears when the electron collision frequency is increased either by increasing 
the gas pressure or removing the Ramsauer minimum in the electron-neutral collision cross section. As Turner et. al. 
\cite{Turner_PRL_76_2069_1996} demonstrated in their paper that increasing the strength of external magnetic field has a 
similar effect as introducing more collisions, i.e. increasing gas pressure, which results in more ohmic heating in the bulk plasma. 
This effect can be seen in figure \ref{fig:figure10} (b) where there is presence of significant ohmic heating in the bulk plasma. Above r = 3.5, 
the increase in total heating is due to the increase in ohmic heating in the bulk region.

\begin{figure}[bht]
\centering
\includegraphics[width=16cm]{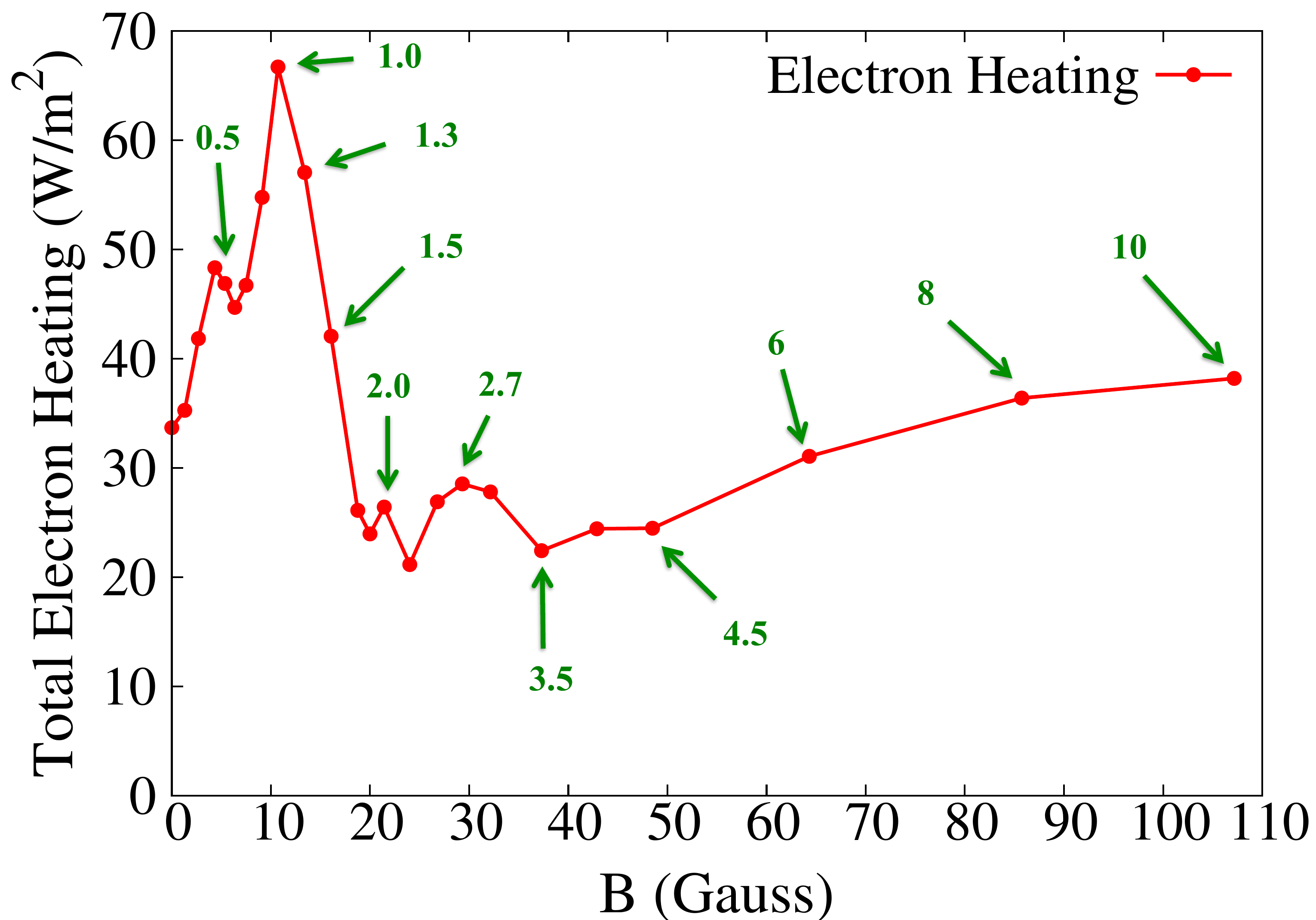}
\caption{Figure shows total electron heating in the discharge for different strength of magnetic field. The values of `r' corresponding to 
different B is also shown in the figure in green font.}
\label{fig:figure11}
\end{figure}

Figure \ref{fig:figure11} shows the variation of total electron heating with strength of magnetic field. The corresponding values of `r' 
are also shown in the graph in green text. This figure indicates that the maximum electron heating is at $\sim$10 G i.e for r = 1 case 
which is $\sim$ 66.7 $W/m^2$. This clearly shows that the total electron heating is maximum at bounce cyclotron resonance 
condition. Increasing values of `r' decreases the total heating and above r = 3.5 it increases again however even at r = 10 the heating 
magnitude is $\sim$ 38 $W/m^2$ which is nearly 43 $\%$ less compared to r = 1 case. As we know that the energy balance is 
between the total heating in the discharge and the energy spent in inelastic collisions (i.e. ionization and excitation) along 
with the kinetic energy lost due to particle fluxes to the electrodes. Thus, the variation in total electron heating is similar to the 
total ionization rate (see figure \ref{fig:figure6}).

\subsection{Electric Field}
\label{electricfield}
\begin{figure}[htp]
\center
\includegraphics[width=16cm,height=20cm]{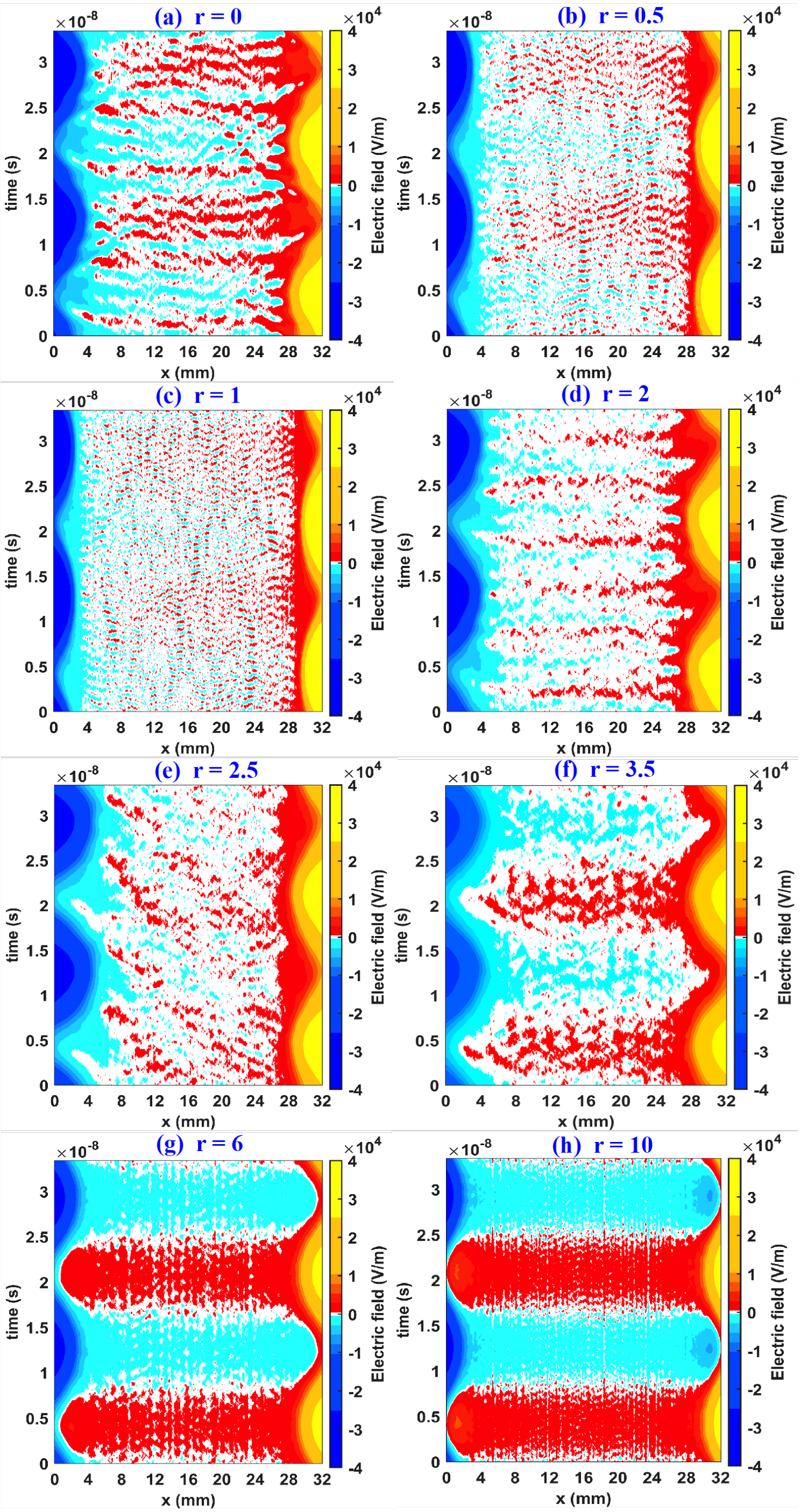}
\caption{Above figure shows the spatiotemporal plots of the electric field for two RF cycles at different magnetic field values. The values of `r' also mentioned in the corresponding sub-plots.}
\label{fig:figure12}
\end{figure}

The electron heating changes from being largely stochastic to largely ohmic when going from unmagnetized/ low magnetic field 
strength to the high magnetic field values. This change is accompanied by an appearance of significant electric field in the bulk, 
as shown in the spatio-temporal plots in figure \ref{fig:figure12}. This figure shows the spatio-temporal profile for two RF cycles, 
which is averaged over last 100 RF cycles after reaching steady state for all simulations. The interesting features such as the 
formation of electric field transients in the bulk plasma for unmagnetized CCP discharges (see figure \ref{fig:figure12} (a) for r = 0) has already been reported in the literature \cite{Sharma_POP_23_2016, Sharma_JPDAP_52_2019, Sharma_POP_26_2019, Sharma_PSST_29_2020, Sharma_JPDAP_54_055205_2021, Sharma_POP_21_2014, Sharma_CPP_55_331_2015}. In this subsection, the effect of external 
magnetic field on these electric field transients will be discussed.

In figure \ref{fig:figure12}, sub-figure (a) for r =0 demonstrates that the sheath electric field penetrates inside the bulk and creates 
filament like structures called electric field transients. As the values of `r' increases to 0.5 ($\sim$ 5.3 G) the density rises to 
$1.0 \times 10^{16}$ $m^{-3}$ from $5.0\times 10^{15}$ $m^{-3}$ i.e. for ‘r=0’ case. Sub-figure (b) shows that the shape 
of transients are finer compared to ‘r=0’ case. When the value of `r' increases from 0 to 0.5 the density of tail end electrons 
or energetic electrons having energy greater than 16 eV also increases from $2.0 \times 10^{14}$ $m^{-3}$ to 
$2.35 \times 10^{14}$ $m^{-3}$, which is nearly an 18 $\%$ increase. When `r' is further increased to 1 (i.e. $\sim$ 10 G), the central 
peak plasma density is nearly $1.7$ $\times $ $10^{16}$ $m^{-3}$. Here the energetic electrons density (energy $>$ 16 eV) is 
also increases to $3.1\times 10^{14}$ $m^{-3}$ which is $\sim$ 32 $\%$ higher compared to r = 0.5 case. The electric field transients 
are further refined and create very fine filament like structures. Such type of filamentation in umagnetized CCP case can be 
seen in the literature where the effect of RF voltage amplitude on electric field transients has been reported \cite{Sharma_JPDAP_52_2019}. 
When r = 2, the bulk peak plasma density decreases to $\sim$ $2.8\times 10^{15}$ $m^{-3}$ and also the density of energetic 
electrons having energy greater than 16 eV i.e. $1.18 \times 10^{14}$ $m^{-3}$ which is nearly 62 $\%$ decrement 
compared to r = 1 case. For this case the transient structures are again thick in nature. Again for further higher values of `r' 
the electrons are confined because of higher magnetic field and significant ohmic heating occurs in the bulk region. The 
thickness of transient structure further increases, filamentation phenomenon of transient electric field vanishes and a lumped 
structure of electric field can be observed in the bulk regime (see sub-figures (g) and (h)). It is to be noted that the transient 
structures for `r = 2.5' are entirely different in nature. This  is currently being investigation and will be reported in a separate publication.

\subsection{Effect of gas pressure on bounce-cyclotron resonance phenomenon}
\label{gaspressure}

As discussed above, the bounce-cyclotron resonance phenomenon occurs at low pressure and in this section we show that by 
increasing pressure the effect of this phenomenon decreases significantly. For that one has to compare the electron-neutral collision frequency with 
the frequency of collisions with the sheath, which is  twice of the cyclotron frequency. The fraction of electrons bouncing off the 
sheath that contributes to the resonant effect is equal to the average probability of an electron NOT colliding with a neutral between 
two bounces. This can be calculated as $exp(-f_{en} \Delta T_{bounce} )=exp(-f_{en}/(2f_{ce} ))$, where $f_{en}$  and $2f_{ce}$ 
are respectively the frequency of electron-neutral collisions and the frequency of bounces. Since $f_{en}$ is proportional to the 
neutral gas pressure, the fraction of electrons contributing to the resonant effect depends on the pressure. When the pressure is 
low enough for this fraction to be significant, the resonant effect is observed. For example, at 5 mTorr in our simulations, on average, 
21$\%$ of the electrons bouncing off the sheath return for a consecutive bounce. For 7.5 mTorr and 10 mTorr, this fraction drops to 
10 $\%$ and 4 $\%$ respectively. Simulations at 10 mTorr were carried out to support this estimate. 

\begin{figure}[htp]
\center
\includegraphics[width=16cm]{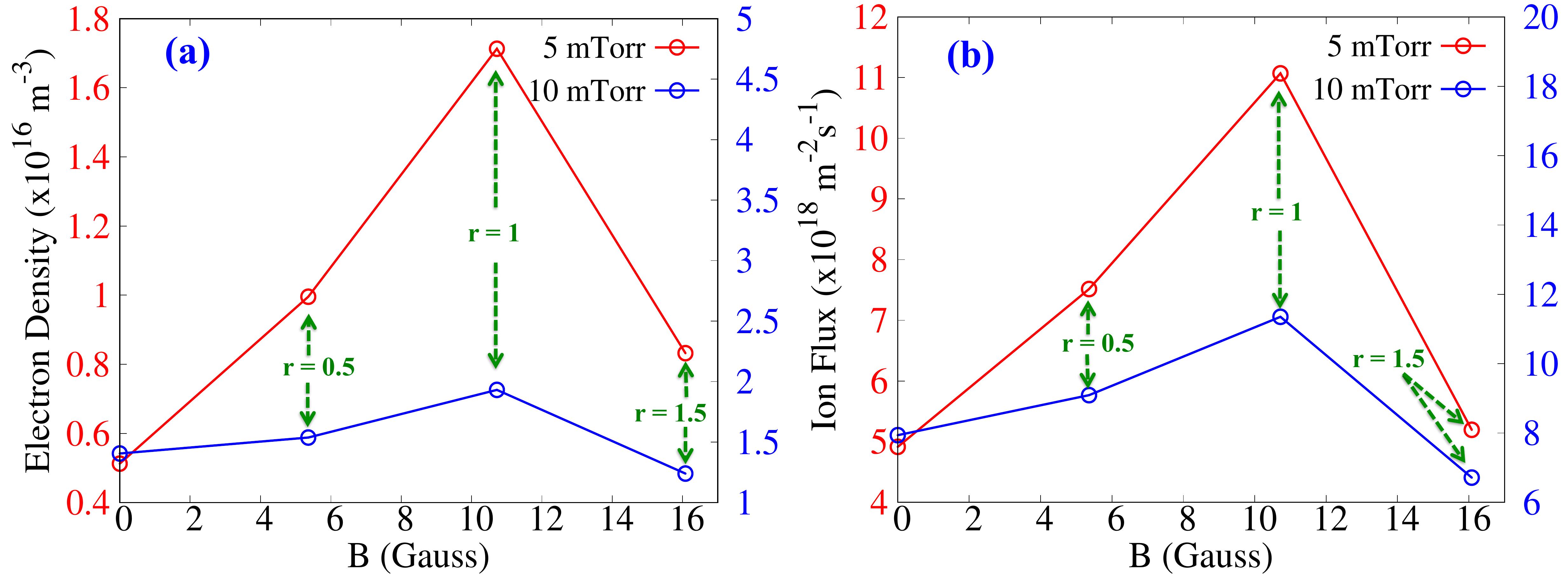}
\caption{Figure shows (a) peak electron density at the centre of discharge and (b) ion flux at the grounded electrode with respect to external magnetic field at 5 mTorr and 10 mTorr case.}
\label{fig:figure13}
\end{figure}

Figure \ref{fig:figure13} (a) and (b) shows the variation of electron density at the center of discharge and ion flux at the grounded 
electrode with respect to the strength of applied magnetic field. In this figure the left y-axis (red font) and right y-axis (blue font) 
indicate the electron bulk density and ion flux for 5 mTorr and 10 mTorr respectively. The bulk electron density and ion flux at the 
grounded electrode for 5 mTorr increases from $\sim$ $5\times 10^{15}$ $m^{-3}$ (at r = 0) to $\sim$ $1.7\times 10^{16}$ $m^{-3}$ 
(at r = 1) and $5\times 10^{18}$ $m^{-2} s^{-1}$ to $11\times 10^{18}$ $m^{-2} s^{-1}$ respectively. Hence the percentage 
increase in electron density and ion flux is nearly 240$\%$ and 120$\%$ respectively. On the other hand for 10 mTorr case, the bulk 
electron density and ion flux at the grounded electrode increases from $\sim$ $1.4\times 10^{16}$ $m^{-3}$ (at r = 0) to 
$\sim$ $1.9\times 10^{16}$ $m^{-3}$ (at r = 1) and $8\times 10^{18}$ $m^{-2} s^{-1}$ to $11.5\times 10^{18}$ $m^{-2} s^{-1}$ respectively. So here the percentage increase in electron density and ion flux is nearly 35$\%$ and 44$\%$ respectively. In conclusion, 
these figures confirm that the resonant effect vanishes at 10 mTorr. Using the above estimate, for the VHF range (30-300 MHz), 
resonant operation would typically require pressure to be less than 10 mTorr.

\section{Conclusion and Discussion}
\label{sumconch4}
\noindent It has been proposed that the resonant condition (r = 1) may be a better operation point for high frequency discharges since 
the nonuniformity for such a low magnetic field is expected to be much smaller compared to high magnetic fields (r$\sim$ 5-10). 
Such a resonant operation is only possible for high frequency discharges as the magnetic field required for cyclotron resonance (r = 1) at 
low RF frequencies will become so low that the Larmor radii of electrons will become larger than the system length, and the effect of 
resonance will disappear. Thus, high frequency CCP discharges have been shown to have a clear advantage through this study.
The performance improvement at higher magnetic fields (r $>$ 3.5) as observed in this study has been consistent with previous studies
\cite{Barnat_PSST_17_2008, Fan_POP_20_2013}. 

It is also shown that at the resonance condition, the density is maximum however the ion energy at the grounded electrode is less compared to the 
unmagnetized case. The total ionization rate is also maximum for the r = 1 case and it is also shown that for this case the ionization inside bulk is 
maximum. For the higher values of `r' ionization rate is maximum near to the sheath edges. With the help of EEDF profile it is demonstrated that 
the tail end electron population is maximum for r = 1 case and the tail depletes for the higher values of `r'. The presence of energetic 
electron population for the different cases of `r' is also figured out with the help of spatiotemporal profile of density of electrons in different energy 
ranges, which shows that the presence of energetic electrons is maximum in the resonance case i.e. for r = 1. 
It is reported that the time average electron heating is maximum for r = 1 case and significant existence of ohmic 
heating is observed for higher values of `r'. The study of electric field transients for different values of `r' has also been reported. Finally 
it is shown that the resonance phenomena is maximum at low pressure i.e. in the collisionless case and it disappears if the neutral gas density 
increases and the plasma is collisional.
Further studies are required for a quantitative comparison between resonant discharges and generally used high magnetic field (50-100 G) 
discharges in terms of the trade-off between ion flux and ion energies at the electrodes.

\section{Acknowledgments}
\noindent 
A.S. is grateful to the Indian National Science Academy (INSA) for the position of an Honorary Scientist.  The work of I.K. was supported by 
the Princeton Collaborative Research Facility (PCRF), which is funded by the U.S. Department of Energy (DOE) under Contract No.
DE-AC02-09CH11466. The results presented in this work have been simulated on ANTYA cluster at Institute for Plasma Research, Gujarat, India.

\section{Author Declarations}
\noindent 
\textbf{Conflict of Interest}
The authors have no conflicts to disclose.

\section{Data Availability}
\noindent 
The data that support the findings of this study are available
from the corresponding author upon reasonable request.

\end{document}